\documentclass[twocolumn,secnumarabic,amssymb, amsmath, nofootinbib,tightenlines, nobibnotes, color, aps, prl,epsfig]{revtex4}
\usepackage{graphicx}
\usepackage{dcolumn}
\usepackage{bm}
\begin{document}
\preprint{APS/123-QED}
\title{ Parametrization of the nuclear structure function }

\author{G.R.Boroun}%
 \email{grboroun@gmail.com; boroun@razi.ac.ir }
 \author{B.Rezaei}%
 \email{brezaei@razi.ac.ir }
\affiliation{ Department of Physics, Razi University, Kermanshah
67149, Iran}

\date{\today}
\begin{abstract}
In this paper, the parametrization of the nuclear structure
function which is directly constrained by the dynamics of QCD in
its high-energy limit is considered. This simple parameterization
of the nuclear structure function is obtained from the proton
experimental data by relying on a Froissart-bounded
parametrization of the proton structure function. This
phenomenological model  describes high-energy QCD in the presence
of saturation effects. Numerical calculations and comparison with
available data from NMC, EMC and E665 Collaborations demonstrate
that the suggested method by N.Armesto, C.A.Salgado and U.A.
Wiedemann (ASW model) provides reliable ratio of nuclear structure
functions $F_{2}^{A}/AF^{p}_{2}$ at low $x$ for light and heavy
nuclei. The magnitude of nuclear shadowing is predicted for
various kinematic regions and can be applied as well in analysis
of ultra-high energy processes by future experiments
at electron-ion colliders. \\
\end{abstract}
 \pacs{***}
\maketitle
\subsection{1. INTRODUCTION}

The knowledge of the QCD dynamics at high energies is essential in
the investigation of hadronic structure studied at current (JLab,
RHIC and the LHC) and future with the Electron-Ion Collider (EIC)
, Large Hadron electron Collider (LHeC), and Future Circular
Collider (FCC) accelerators on the horizon. One of the main goals
of high energy nuclear physics is to comprehend the sub structure
of nucleons in the framework of QCD which is a successful theory
in describing the hadronic and nuclear phenomena as well as the
inner structure of nucleon and nuclei. In this regard, the
structure functions of nucleon and nuclei have played a crucial
role. Nuclear structure functions measured in deep inelastic
scattering (DIS) experiments (which have been performed by NMC,
SLAC, NMC, FNAL, BCDMS, HERMES, and JLAB groups) offer valuable
information for understanding the dynamics of partons in the
nuclear environment. The correct characterisation of nuclear
effects in the parton distribution functions (PDFs) is important
due to their relevance in the determination of the proton PDFs and
this is baseline for new phenomena in heavy-ion collisions.\\
Nuclear parton distributions (nPDF) are needed in the computation
of inclusive cross sections of hard, factorizable, processes in
high energy nuclear collisions. The nuclear effects are generally
added as a modification of a baseline PDFs, either by expanding
the parametrization or multiplying it by a factor as reported in
Refs.[1-12]. There is, in both cases, a dependence on the atomic
mass number A . For this reason, there is a continual need for new
data sets to broaden global analyses. In this way, groups like
EPPS [13], NNPDF [14] or nCTEQ [15] have been demonstrated the
nucleon and nuclear PDF (nPDF) analyses in recent years. Nuclear
effects on parton distributions and structure functions are
important for interpreting high energy processes involving nuclei
such as heavy ions and electron-nucleus collisions at EIC [16] and
LHeC/FCC [17]. These colliders (i.e., EIC and LHeC/FCC) are
constructive in understanding the momentum distribution of quarks
and gluons in nuclei.\\
At small values of the Bjorken variable $x$, the non-linear QCD
effects will  consider in these colliders related to the studies
of partonic structure of protons and nuclei [18, 19]. The proposed
LHeC collider covers a wide kinematical range down to
$x{\sim}10^{-6}$ in the perturbative range
$Q^{2}{\gtrsim}1\mathrm{GeV}^{2}$ making it an ideal machine to
study small-x physics. In addition to the Large Hadron-electron
Collider, the construction of an Electron-Ion Collider with a
possibility to operate with a wide variety of nuclei, will allow
one to explore the low-$x$ region in much greater detail. In this
region a transition between linear and non-linear scale evolution
of the parton densities will be crucial [20]. The latter regime,
known as "saturation" [21, 22], occurs at low $x$ and low
interaction scale $Q^{2}$ where the recombination of low
$x$ gluons becomes increasingly important.\\
Also the small $x$ region of QCD can be described theoretically in
the effective field theory known as the Color Glass Condensate
(CGC); see Ref. [23] for a review. In the CGC picture the
non-linear evolution equations describe the evolution of the small
$x$ gluon fields. Probing these nonlinearities at the LHeC and EIC
are crucial to test the saturation picture [18]. These
nonlinearities are important in electron-nucleus  scattering in
comparison with the electron-proton interactions. The Hadron
Electron Ring Accelerator (HERA) did not report the non-linear
behavior of the distribution functions in deep inelastic
scattering off nuclei at collider energies. Despite this, the
future colliders (such as LHeC and EIC) will offer unique
capabilities to answer the non-linearity in
nuclei at high energies [20,24].\\
The proposed EIC would enable the first direct measurements of
nuclear gluons at intermediate and large $x$ using heavy quark
probes and could qualitatively advance our understanding of the
gluonic structure of nuclei [25]. This  collider (i.e., EIC) would
have a strong impact, in particular on understanding the small-
and large-$x$ regions of nuclear shadowing and the EMC effect in
comparison with fixed-target kinematics, which DIS data
considerably restricts their range in $x$ and $Q^{2}$, and only
with limited statistics for various nuclei [26]. The experiment at
EIC is DIS off a proton or a nucleus with the variable
center-of-mass energy within the range
$20<\sqrt{s}<140~\mathrm{GeV}$, where this is lower than at HERA
with $\sqrt{s}=318~\mathrm{GeV}$, but the luminosity is higher by
a factor of 1000. The EIC will combine the experience from HERA to
deliver polarized electron beams with the experience from RHIC to
be the first machine that provides the collision of polarized
electrons with polarized protons, and at a later stage, polarized
$^{2}\mathrm{H}$ and $^{3}\mathrm{He}$ [27]. At fixed-target
facilities, such as JLab, the majority of the momentum is carried
by the electron, while for electron-ion collider experiments, the
majority of the momentum is carried by the ion beam, so variables
are defined according to the electron beam and the ion beam
(against the electron beam), respectively. A detailed description
of the fixed-target and the EIC 4-momenta is given in [28]. For
collider experiments, the center-of-mass energy
$\sqrt{s_{\mathrm{EIC}}}=\sqrt{4E_{e}E_{h}}$ is often used as a
frame of reference and for fixed target experiments the
center-of-mass energy
$\sqrt{s_{\mathrm{JLab}}}=\sqrt{2m_{h}E_{e}+m^2_{h}}$ where
$E_{h}=m_{h}$ (target mass). The familiar definitions of
Bjorken-$x$ are significantly different between EIC and
fixed-target experiments, as for fixed-target experiments
$x_{\mathrm{Fixed}}=Q^2/(2m_{h}\nu)$ where $\nu$ is the energy
transform $\nu=E_{e}-E'_{e}$. In JLab the Bjorken-$x$ scaling is
defined $x_{\mathrm{JLab}}=Nx_{\mathrm{Fixed}}$, where $N$ is the
number of nucleons in the target. In collider experiments
$x=\frac{x_{p}}{N}$ where $x_{p}=Q^2/[2E_{p}(\nu+\nu_{z})]$ and
$\nu_{z}=E_{e}-E'_{e}\cos{\theta_{e}}$. Also the collider
definition of $W^2$ is defined $W^2=Q^2(1-x)/x$ where in
fixed-target experiments it is modified by
$W_{\mathrm{JLab}}^2=m_{p}^2+Q^2(1-x_{\mathrm{JLab}})/x_{\mathrm{JLab}}$.
Therefore, the kinematic ($x$,$Q^2$) range of the fixed-target
experiments  will be test by EIC as discussed in Ref.[28].\\
The simplest observable to study nuclear effects is to measure the
structure function ratios at small $x$ ($x{\leq}0.01$, shadowing
region). Indeed, the structure function $F_{2}$ per nucleon turns
out to be smaller in nuclei than in a free nucleon [24] and this
is very important for the study of nuclear structure and nuclear
collisions. Nuclear shadowing is a consequence of multiple
scattering, as this is well understood in the gluon recombination.
Indeed in the frame in which the nucleus is moving fast, the gluon
clouds from different nucleons overlap. Therefore the ratio of the
structure functions (i.e., $F_{2}^{A}/AF_{2}^{p}$), at small $x$,
is smaller than 1. The shadowing effect is well understood by the
characteristic momentum scale which is known as the saturation
scale $Q^{2}_{s}$. This scale (i.e., saturation scale) increases
with decreasing $x$ as $Q^{2}_{s}=Q_{0}^{2}(x/x_{0})^{-\lambda}$.
Geometrical scaling for $\alpha_{s}(Q^{2})xg(x,Q^{2})/Q^{2}$ holds
at the boundary $Q^{2}=Q^{2}_{s}$. For $Q^{2}<Q^{2}_{s}$ the
linear evolution is strongly perturbed by nonlinear effects and
for $Q^{2}>Q^{2}_{s}$
the nonlinear screening effects can be neglected [29].\\
In this paper a simple parametrization for the nuclear structure
function based on the parametrization of the proton structure
function is proposed. In Ref.[30] authors have suggested
parametrization of the proton structure function which describes
fairly well the available experimental data on the reduced cross
sections at small $x$ where it is also pertinent in investigations
of lepton-hadron processes at ultra-high energies(i.e., the
scattering of cosmic neutrinos from hadrons). The parametrization
of the proton structure function describes all data on DIS in the
region of $x {\leq}0.01$ in a wide interval of photon virtualities
[30]. Relying on saturation scaling arguments, a simple model for
the parameterization of the
nuclear structure function is suggested.\\

\subsection{2. A MODEL FOR THE NUCLEAR STRUCTURE FUNCTION}

It is customary to write the proton structure function $F_{2}$
into the cross sections $\sigma_{T,L}$ for the collision of the
transversal ($T$) or longitudinal ($L$) virtual photon of momentum
$q$, $q^{2}=-Q^{2}$, on the proton as follows
\begin{eqnarray}
F^{p}_{2}(x,Q^{2})&=&\frac{Q^{2}}{4{\pi^{2}}\alpha_{em}}[\sigma^{\gamma^{*}p}_{T}(x,Q^{2})+\sigma^{\gamma^{*}p}_{L}(x,Q^{2})].
\end{eqnarray}
The electron-proton (ep) deep inelastic scattering (DIS) data at
small values of the Bjorken variable $x$ can be described within
the framework of the dipole model [24,31-35]. In the dipole frame,
the incoming photon splits into a $q\overline{q}$ which then
interacts with the proton. This process depends on the total
dipole-proton cross section, which varies with $x$ and the
transverse size $r$ of the dipole. The total $\gamma^{*}p$ cross
section reads
\begin{eqnarray}
\sigma^{\gamma^{*}p}_{L,T}(Q^{2},Y)=\int d^{2}\mathbf{r}
\int_{0}^{1}dz|\Psi_{L,T}(\mathbf{r},z;Q^{2})|^{2}\sigma_{dip}^{\gamma^{*}p}(r,Y),
\end{eqnarray}
with $Y=\log(1/x)$ called the rapidity. $\Psi_{L,T}$ is the wave
function for the splitting of the virtual photon into a
$q\overline{q}$ pair (dipole) and
$\sigma_{dip}^{\gamma^{*}p}(r,Y)=2{\pi}R_{p}^{2}N(r,Y)$ is the
dipole-proton cross section where $N$ is the dipole-proton
scattering amplitude as entering the QCD evolution equations. Here
$z$ is a fraction of longitudinal photon momentum carried by quark
and $R_{p}$ is the radius of the proton. The wave function of the
virtual photon, $|\Psi|^{2}=|\Psi_{T}|^{2}+|\Psi_{L}|^{2}$, in the
leading order is given by
\begin{eqnarray}
|\Psi_{T}(r,z;Q^{2})|^{2}&=&\frac{3\alpha_{em}}{2\pi^{2}}\sum_{f}e_{f}^{2}
\{[z^{2}+(1-z)^{2}]\overline{Q}_{f}^{2}K^{2}_{1}(r\overline{Q}_{f})\nonumber\\
&&+m_{f}^{2}
K^{2}_{0}(r\overline{Q}_{f}) \},\nonumber\\
|\Psi_{L}(r,z;Q^{2})|^{2}&=&\frac{3\alpha_{em}}{2\pi^{2}}\sum_{f}e_{f}^{2}
4Q^{2}z^{2}(1-z)^{2}K^{0}_{1}(r\overline{Q}_{f}),
\end{eqnarray}
where $K_{0,1}$ are the Bessel functions, sums $\sum_{f}$ runs
over all quark flavors with charge $e_{f}$ and mass $m_{f}$ and
also $\overline{Q}_{f}^{2}=z(1-z)Q^{2}+m_{f}^{2}$.\\
In the color dipole formalism the nuclear structure function
(i.e., $F_{2}^{A}$) is proportional to the dipole nucleus cross
section (i.e., $\sigma_{dip}^{\gamma^{*}A}$).
$\sigma_{dip}^{\gamma^{*}A}$ describes the interaction of the
$q\overline{q}$ dipole with the nucleus target. In the eikonal
approximation, the total cross-section for dipole to scatter off
the target nucleus at an impact parameter $\mathbf{b}$ is given by
[1-4,19-35]
\begin{eqnarray}
\sigma_{dip}^{\gamma^{*}A}(r,Y)&=&2\int
d^{2}\mathbf{b}N_{A}(\mathbf{r},Y;\mathbf{b}).
\end{eqnarray}
The nuclear scattering amplitude $N_{A}(\mathbf{r},Y;\mathbf{b})$
dependents upon the impact parameter $\mathbf{b}$, rapidity and
dipole size $\mathbf{r}$.\\
The one dimensionless variable $\tau=Q^{2}/Q^{2}_{s}$ is the
geometric scaling where the physics remains unchanged when one
moves parallel to the saturation line. Indeed the saturation scale
is a border between dense and dilute gluonic systems and the
geometric scaling can be understood as a property of the small $x$
evolution equations in the large rapidity regime. In Ref. [36], an
analytic interpolation of lepton-proton data as function of the
scaling variable $\tau=Q^{2}/Q^{2}_{s}$ was proposed. The ASW form
of the single universal curve on $\sigma^{\gamma^{*}p}$ is given
by [35]
\begin{eqnarray}
\sigma^{\gamma^{*}p}(x,Q^{2}){\equiv}\Phi(\tau)=
\overline{\sigma}_{0}[\gamma_{E}+\Gamma(0,\xi)+{\ln}\xi]
\end{eqnarray}
where $\gamma_{E}$ and $\Gamma(0,\xi)$ are the Euler constant and
the incomplete $\Gamma$ function respectively. Authors in Ref.[35]
extracted the $\xi$ function from a fit to lepton-proton data as
$\xi=a/\tau^{b}$ with $a=1.868$ and $b=0.746$. The normalization
is fixed by $\overline{\sigma}_{0}=40.56~\mu{b}$ and the
saturation scale $Q^{2}_{s}$ is parametrized as
$Q^{2}_{s}=1~\mathrm{GeV}^{2}(\overline{x}/x_{0})^{-\lambda}$,
where $x_{0}=3.04{\times}10^{-4}$, $\lambda=0.288$ and
$\overline{x}=x(Q^{2}+4m^{2}_{f})/Q^{2}$ with
$m_{f}=0.14~\mathrm{GeV}$. The nuclear structure function is
defined by
$F_{2}^{A}(x,Q^{2})=Q^{2}\sigma^{\gamma^{*}A}/(4\pi^{2}\alpha)$
where
$\sigma^{\gamma^{*}A}=\frac{{\pi}R^{2}_{A}}{{\pi}R^{2}_{p}}\sigma^{\gamma^{*}p}(\tau_{A})$
and $R_{A}$ is the nuclear radius.\\
In Ref.[33] authors have introduced the saturation scale
$Q_{s}^{2}(Y){\propto}\exp(\upsilon_{c}Y)$ where it is based on an
analytic interpolation  asymptotic behavior of the amplitude for
the unintegrated gluon function. The dipole-proton (nucleus)
scattering amplitude is as a result of the Balitsky-Kovchegov (BK)
evolution equation [37-38] at high-energy evolution.
$Q_{s}^{2}(Y)$ is obtained from the knowledge of the
Balitsky-Fadin-Kuraev- Lipatov (BFKL) kernel [39] in the form [33]
\begin{eqnarray}
Q_{s}^{2}(Y)&=&k_{0}^{2}\exp(\overline{\alpha}\upsilon_{c}Y),
\end{eqnarray}
where the parameters of model have been determined according to
the HERA data [40] as $\upsilon_{c}=0.807$,
$k_{0}^{2}=3.917{\times}10^{-3}~\mathrm{GeV}^{2}$ and
$\overline{\alpha}=3\alpha_{s}/\pi$. The condition for geometric
scaling in the case of $\gamma^{*}-A$ interactions is defined by
the following form [33,35]
\begin{eqnarray}
\sigma_{tot}^{\gamma^{*}A}\bigg(\frac{Q^{2}}{Q_{s,A}^{2}}\bigg)&=&
\bigg(\frac{\pi{R_{A}^{2}}}{\pi{R_{p}^{2}}}\bigg){\times}
\sigma_{tot}^{\gamma^{*}p}\bigg(\frac{Q^{2}}{Q_{s,A}^{2}}\bigg),
\end{eqnarray}
where
\begin{eqnarray}
Q_{s,A}^{2}(Y)&=&Q_{s,p}^{2}(Y)\bigg(
\frac{A{\pi}R_{p}^{2}}{{\pi}R_{A}^{2}}\bigg)^{1/\delta},
\end{eqnarray}
and $Q_{s,p}(Y)({\equiv}Q_{s}(Y))$ is the saturation scale for a
proton target. In the dipole model, the nuclear data are
reproduced for $\delta=0.79{\pm}0.02$ and
${\pi}R_{p}^{2}=1.55{\pm}0.02~\mathrm{fm}^{2}$ at low values of
$x$ where the nuclear radius is given by the usual parametrization
$R_{A}=(1.12A^{1/3}-0.86A^{-1/3})\mathrm{fm}$ [33,35,41-42]. In
the region of small $x$ , the effect of nuclear shadowing
manifests itself as an inequality $F_{2}^{A}/AF_{2}^{p}$ where
$\sigma^{\gamma^{*}A}/A\sigma^{\gamma^{*}p}{\approx}F_{2}^{A}/AF_{2}^{p}$.
This leads to the following result for the nuclear structure
function
\begin{eqnarray}
F_{2}^{A}(\tau_{A},Y)&=&
\bigg(\frac{\pi{R_{A}^{2}}}{\pi{R_{p}^{2}}}\bigg){\times}
F_{2}^{p}(\tau_{A},Y),
\end{eqnarray}
where the  geometric scaling is considered by the following form
\begin{eqnarray}
\tau_{A}&=&\tau_{p}\bigg(
\frac{{\pi}R_{A}^{2}}{A{\pi}R_{p}^{2}}\bigg)^{1/\delta},
\end{eqnarray}
and $\tau_{p}{\equiv}\tau=Q^{2}/Q^{2}_{s}(Y)$. The explicit
expression for the parametrization of the nuclear structure
function $F_{2}^{A}$, is the same  of the parametrization of the
proton structure function with the change
$Q^{2}{\rightarrow}Q^{2}_{A}$, where
\begin{eqnarray}
Q^{2}_{A}&=&\tau\bigg(\frac{A{\pi}R_{p}^{2}}{
{\pi}R_{A}^{2}}\bigg)^{1/\delta}Q_{s}^{2}\nonumber\\
&&=\tau Q^{2}_{s,A}.
\end{eqnarray}
In Ref.[30] an analytical expression for the proton structure
function, which describes fairly well the available experimental
data on the reduced cross section in full accordance with the
Froissart predictions, is defined by the following form
\begin{eqnarray}
F_{ 2}(x,Q^{2})& =& D(Q^{2})(1-
x)^{n}\sum_{m=0}^{2}A_{m}(Q^{2})L^{m}.
\end{eqnarray}
Therefore the parametrization of the nuclear structure function
(according to the Eqs.(9-12)) reads as
\begin{eqnarray}
F_{2}^{A}(\tau,Y)&=&\bigg(\frac{\pi{R_{A}^{2}}}{\pi{R_{p}^{2}}}\bigg)D(\tau Q^{2}_{s,A})(1-e^{-Y})^{n}{\times}\nonumber\\
&&\sum_{m=0}^{2}A_{m}(\tau Q^{2}_{s,A})L^{m}(\tau Q^{2}_{s,A},Y),
\end{eqnarray}
where
\begin{eqnarray}
D(Q^{2}_{j})&=&\frac{Q^{2}_{j}(Q^{2}_{j}+\lambda
M^{2})}{(Q^{2}_{j}+M^{2})^{2}},\nonumber\\
A_{0}(Q^{2}_{j})&=&a_{00}+a_{01}L_{2}(Q^{2}_{j}),\nonumber\\
A_{i}(Q^{2}_{j})&=&\sum_{k=0}^{2}a_{ik}L_{2}(Q^{2}_{j})^{k},~~i=(1,2),\nonumber\\
L(Q^{2}_{j},Y)&=&Y+{\ln}\frac{Q^{2}_{j}}{Q^{2}_{j}+\mu^{2}},\nonumber\\
L_{2}(Q^{2}_{j})&=&{\ln}\frac{Q^{2}_{j}+\mu^{2}}{\mu^{2}},~~~~~~~j=p,A.
\end{eqnarray}
The effective parameters are defined in Table I.\\
In the ASW model [35], the nucleon structure function is defined
by the following form as
\begin{eqnarray}
F_{2}^{p}(\tau)&=&\frac{Q_{s,p}^{2}\tau_{p}}{4\pi^{2}\alpha}[\gamma_{E}+\Gamma(0,\xi_{p})+{\ln}\xi_{p}].
\end{eqnarray}
The explicit expression for the ratio in this model reads as
\begin{eqnarray}
\frac{F_{2}^{A}(\tau_{A})}{AF_{2}^{p}(\tau)}=\bigg(\frac{\pi{R_{A}^{2}}}{\pi{R_{p}^{2}}}\frac{Q_{s,A}^{2}\tau_{A}}{AQ_{s,p}^{2}\tau_{p}}\bigg)
\frac{\gamma_{E}+\Gamma(0,\xi_{A})+{\ln}\xi_{A}}{\gamma_{E}+\Gamma(0,\xi)+{\ln}\xi},
\end{eqnarray}
where $\xi_{A}=a/\tau_{A}^{b}$. Notice that Eq.(15) is written in
the geometrical scaling whereas the equation (12) is in
($x-Q^{2}$) space. Since the cross section in Eq.(7) only depends
on $Q^{2}/Q^{2}_{s}$, the replacement
$Q^{2}_{s,p}{\rightarrow}Q^{2}_{s,A}$ corresponds to the rescaling
$Q^{2}{\rightarrow}Q^{2}/\lambda_{A}^{2}$ in Eq.(12), where
$\lambda_{A}=(\frac{\pi R_{A}^{2}}{A\pi R_{p}^{2}})^{1/2\delta}$ [43].\\

\subsection{3. RESULTS AND CONCLUSIONS}

In this section, the numerical calculation of the nuclear
structure function and the nuclear ratio using Eqs. (13) and (14)
is investigated. With respect to these equations the nucleon and
nuclear structure functions and the corresponding ratios for
values $x{\leq}0.01$ can be computed. Calculations have been
performed at a fixed value of the running coupling. For the LO
BFKL kernel, one finds $\upsilon_{c}=0.807$ and
$\upsilon_{c}=4.88\overline{\alpha}$ [33], therefore the coupling
constant is fixed by $\alpha_{s}=0.17$. The overlap between the
models indicates that the Bjorken variable $x$ to vary in the
interval $10^{-2}{\leq}x{\leq}10^{-6}$ and $Q^{2}$ varies in the
interval
$0.15~\mathrm{GeV}^{2}{\leq}Q^{2}{\leq}150~\mathrm{GeV}^{2}$.
\begin{figure}[h]
\includegraphics[width=0.45\textwidth]{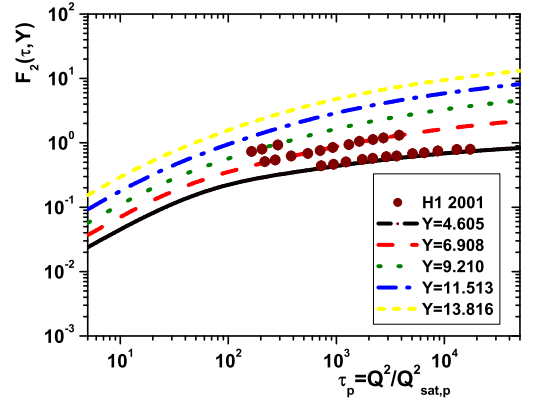}
\caption{The proton structure function $F_{2}(\tau,Y)$ plotted
versus scaling variable $\tau_{p}=Q^{2}/Q^{2}_{\mathrm{sat,p}}$
for different values of rapidities $Y=\ln\frac{1}{x}$ from
$Y_{\mathrm{min}}=4.605$ (solid curve) to
$Y_{\mathrm{max}}=13.816$ (short-dash curve) compared with the H1
Collaboration data as accompanied with total errors [40], for
$x=10^{-6},...,10^{-2}$ (curves from up to down, respectively).
}\label{Fig1}
\end{figure}
 In
Fig.1 the proton structure functions are presented as a function
of scaling variable $\tau$ for different values of rapidities and
compared with the H1 Collaboration data [40]. The proton structure
functions obtained into the scaling variable $\tau$ are comparable
with data of the H1-Collaboration. At intermediate and high
scaling variable $\tau$ the extracted proton structure functions
are in a good agreement with experimental data. The
parametrization of the proton structure function is translated to
the nuclear structure function as quantified by Eq.(13). To
investigate the parametrization model for the nuclear structure
function we shall consider results obtained using the scaling
variable $\tau$ for light and heavy nuclei.
\begin{figure}[h]
\includegraphics[width=0.45\textwidth]{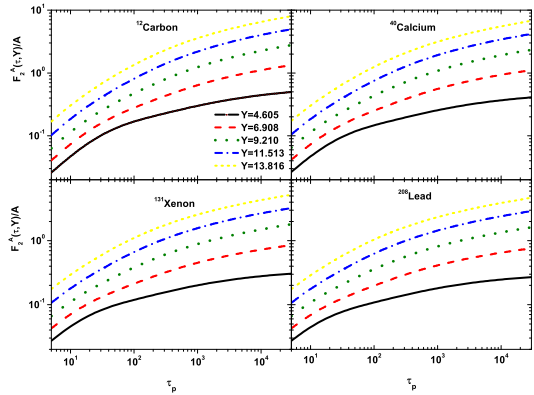}
\caption{  The nuclear structure function $F^{A}_{2}(\tau,Y)$
(normalized to the number of nucleons) plotted versus scaling
variable $\tau_{p}=Q^{2}/Q^{2}_{\mathrm{sat,p}}$ for different
values of rapidities $Y=\ln\frac{1}{x}$, from
$Y_{\mathrm{min}}=4.605$ (solid curve) to
$Y_{\mathrm{max}}=13.816$ (short-dash curve) for Carbon, Calcium,
Xenon and Lead nuclei, for $x=10^{-6},...,10^{-2}$ (curves from up
to down, respectively). }\label{Fig2}
\end{figure}
In Fig.2 we show the nuclear structure function  as a function of
scaling variable $\tau$ for different values of rapidities and
different nuclei. These theoretical curves are the result of the
parametrization of the nuclear structure function, meaning that we
are using the geometrical scaling in the parametrization method.
Furthermore, we check the $A$-dependence of the nuclear structure
function at $Y=9.210$ in Fig.3.
\begin{figure}[h]
\includegraphics[width=0.45\textwidth]{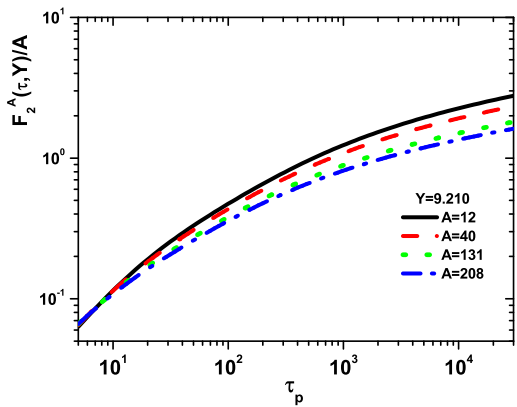}
\caption{$A$-dependence of nuclear structure function (normalized
to the number of nucleons) for $A=12, 40, 131$ and $208$ at
$Y=9.210~ (x=10^{-4})$ plotted versus scaling variable $\tau_{p}$
. }\label{Fig3}
\end{figure}
 In this figure (i.e., Fig.3), the
nuclear structure function decreases as the atomic mass number $A$
increases. It is clear that the $A$- dependence of the nuclear
structure function in electron-nucleus collisions at mid-rapidity
involve additional nuclear effects
which are at least as significant as nuclear shadowing.\\
The parameterization method reported by authors in Ref.[35] can be
covered all data on photon-nucleon and photon-nuclei, so we
compared the nuclear ratio $F_{2}^{A}/AF_{2}^{p}$  with the
experimental data [44] in Fig.5. Figure 5 compares our
calculations of the shadowing due to the ASW model (Eq.(16)) with
available data from the E665, EMC  and EMC Collaborations [44].
The shadowing in nuclei is studied in this figure (i.e., Fig.4)
through the ratios of cross sections per nucleon for different
nuclei at the Bjorken scaling values $x=10^{-1},..,10^{-3}$
respectively. We have selected data where $x<0.03$ and accompanied
with total errors. We observe that for fixed $x$ and large
$Q^{2}$, the ratio become closer to one, i.e shadowing decreases
with increasing $Q^{2}$ and also a larger nuclear shadowing is
visible for Lead target at low $Q^{2}$. These results are
comparable with others in Refs.[9,42]. In Refs.[9] and [42],
nuclear shadowing in the Regge limit within the Glauber-Gribov
model and in the color dipole formalism based on the rigorous
Green function techniques at small $x$ have been considered
respectively. The behaviors and magnitudes of shadowing using the
both color dipole formalism from the higher $|q\overline{q}>$ Fock
component in Ref.[42] and the parametrization method are
comparable. The predictions for expected scaling kinematics in
experiments at EICs are presented in Fig.4 for  the C, Ca, Xe and
Pb targets. Also the ASW model is in good agreement with the
dipole model calculation of [45] where rescatterings of the full
$q\overline{q}+Ng$ fluctuation is taken into account where the
higher Fock-states of the dipole correspond to the summation of
triple-pomeron diagrams in that approach. Therefore, these
predictions within the parametrization of the nuclear ratios, due
to the ASW model, are comparable with other dipole models (such as
GBW[46-47], KST[48],
BGBK[49], IP-sat[50]).\\
\begin{figure}[h]
\includegraphics[width=0.55\textwidth]{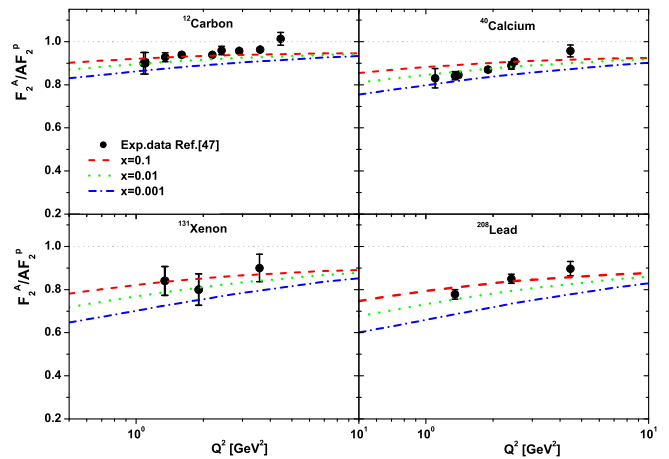}
\caption{The nuclear ratio $F_{2}^{A}/AF_{2}^{p}$ with respect to
the ASW model, as a function of $Q^{2}$ for several nuclear
targets for $x=10^{-1},...,10^{-3}$ (curves from up to down,
respectively), compared with the experimental data [44] (i.e.,
E665, EMC and NMC Collaborations) as accompanied with total
errors. }\label{Fig5}
\end{figure}
In conclusion, we studied the shadowing in deep-inelastic
scattering off nuclei in the kinematic regions accessible by the
future electron-ion colliders with respect to the parametrization
method and the ASW model respectively. We presented a further
development of the parametrization of the DIS structure function
with respect to the saturation scaling. We calculated the
shadowing of the nuclear structure function in the ASW model in
the region of $x{\leq}0.1$ in a wide interval of photon
virtualities. Then we compared the magnitudes of shadowing using
the ASW  model for the light and heavy nuclei, and showed that
that these predictions are in a good agreement with available data
from the E665, EMC and NMC
collaborations.\\

\subsection{ACKNOWLEDGMENTS}

The authors are thankful to the Razi University for financial
support of this project.
 Also G.R.Boroun wishes to especially thank N.Armesto for reading and commenting on the manuscript.\\

\begin{table}[h]
\caption{ The effective parameters at low $x$ are defined by the
Block-Halzen fit to the HERA data as $M^{2}=0.753 \pm 0.068~
\mathrm{GeV}^{2}$, $\mu^2 = 2.82 \pm 0.290~ \mathrm{GeV}^{2}$,
$n=11.49\pm 0.99$ and $\lambda= 2.430~\pm 0.153$  [30].}
\begin{tabular} {cccc}
\toprule \\  \multicolumn{2}{c}{parameters \quad \quad \quad ~~~~~~~~~~~~~~~~value}    \\ &&&\\ \hline \\ &&&\\
  $a_{10} $  &   \quad  $8.205\times 10^{-4}~~  \pm  4.62\times10^{-4} $  \\

  $a_{11} $  &   \quad   $-5.148\times 10^{-2}\pm 8.19\times10^{-3}$  \\

  $a_{12}$   &    \quad  $-4.725\times 10^{-3}\pm 1.01\times10^{-3}$   \\  &&&\\

 $a_{20}$   &   \quad   $2.217\times 10^{-3}\pm 1.42\times10^{-4} $ \\

 $a_{21}$   &   \quad   $1.244\times 10^{-2}\pm 8.56\times10^{-4}$  \\

 $a_{22}$    &    \quad  $5.958\times 10^{-4}\pm 2.32\times10^{-4} $ \\ &&& \\

$a_{00}$& \quad  $2.550\times 10^{-1}~\pm 1.600\times10^{-2}$ & &\\

$a_{01}$& \quad  $1.475\times 10^{-1}~\pm 3.025\times10^{-2}$ & &\\

\hline

\end{tabular}
\end{table}
\section{References}
1. B.Nachman, K.Wichmann and P.Zurita, arXiv [hep-ph]:
2108.09849.\\
2. N.Armesto, H.Paukkunen, C.A.Salgado and K.Tywoniuk, Phys.Lett.B {\bf694}, 38 (2010).\\
3. K.J.Eskola, H.Honkanen, V.J.Kolhinen and C.A.Salgado,
Phys.Lett.B {\bf532}, 222 (2002); arXiv
[hep-ph]:0302170.\\
4. K.J.Eskola, H.Paukkunen and C.A.Salgado, JHEP {\bf0807}, 102 (2008); JHEP {\bf0904}, 065 (2009).\\
5. S.Kumano and M.Miyama, Phys.Lett.B {\bf378}, 267 (1996); M.Arneodo, Phys.Rept.
{\bf240}, 301 (1994); D.F. Geesaman, K.Saito and A.W.Thomas,
Ann.Rev.Nucl.Part.Sci. {\bf45}, 337 (1995).\\
6. E.P.Segarra et al., Phys.Rev.D {\bf103}, 114015 (2021).\\
7. S.Heidari, B.Rezaei, J.K.Sarma and G.R.Boroun, Nucl.Phys.A
 {\bf986}, 195 (2019); S.Heidari, B.Rezaei  and G.R.Boroun, Int.J.Mod.Phys.E {\bf26},
1750067 (2017); G.R.Boroun,  B.Rezaei and S.Heidari,
Int.J.Mod.Phys.A {\bf32}, 1750197 (2017).\\
8. S.Atashbar Tehrani, Phys.Rev.C {\bf86}, 064301 (2012);
J.Sheibani, A.Mirjalili and S.Atashbar Tehrani, Phys.Rev.C
{\bf98}, 045211 (2018) ; H.Khanpour, M.Soleymaninia, S.Atashbar
Tehrani, H.Spiesberger and  V.Guzey, Phys.Rev.D {\bf104}, 034010 (2021).\\
9. N.Armesto, A.B.Kaidalov, C.A.Salgado and K.Tywoniuk, Eur.Phys.J.C {\bf68}, 447 (2010).\\
10. F.Zaidi et al., Phys.Rev.D {\bf99}, 093011 (2019).\\
11. J.L.Albacete et al., CERN-TH/2003-184, arXiv [hep-ph]:0308050 ; N.Armesto et al., Eur.Phys.J.C {\bf29}, 531 (2003).\\
12. S.J.Brodsky, V.E.Lyubovitskij and I.Schmidt, SLAC-PUB-17626, arXiv [hep-ph]:2110.13682.\\
13. K.J.Eskola, P.Paakkinen, H.Paukkunen and C.A.Salgado, Eur.
Phys.J.C {\bf77}, 163 (2017).\\
14. R.D.Ball et al., Eur.Phys.J.C {\bf77}, 663 (2017).\\
15. A.Accardi, T.J.Hobbs, X.Jing and P.M.Nadolsky, Eur.Phys.J.C
{\bf81}, 603 (2021).\\
16. A.Accardi, et al., arXiv:1212.1701; D.P.Anderle et al.,
Front.Phys.{\bf16}, 64701 (2021).\\
17. LHeC Collaboration and FCC-he Study Group, P. Agostini et al., J. Phys. G: Nucl. Part. Phys. {\bf48}, 110501(2021).\\
18. Heikki M$\ddot{\mathrm{a}}$ntysaari, arXiv [hep-ph]: 1811.06328.\\
19. V.P.Goncalves, EPJ Web of Conferences {\bf112}, 02006
(2016).\\
20. E.C. Aschenauer et al., Phys.Rev.D {\bf96}, 114005 (2017).\\
21. J.Jalilian-Marian and Y.V.Kovchegov, Prog. Part. Nucl. Phys.
{\bf56}, 104 (2006).\\
22. J. L. Albacete and C. Marquet, Prog. Part. Nucl.Phys. {\bf76},
1 (2014).\\
23. F.Gelis, E.Iancu, J.Jalilian-Marian and R. Venugopalan,
Annu.Rev.Nucl.Part.Sci.{\bf60}, 463 (2010); L. D. McLerran and R.
Venugopalan, Phys.Rev.D {\bf49}, 3352 (1994);  E. Iancu, A.
Leonidov and L. D. McLerran, Nucl. Phys. A {\bf692},  583
(2001).\\
24. N. Armesto, J.Phys.G {\bf32}, R367 (2006); E.Sichtermann, Nucl.Phys.A {\bf956}, 233 (2016)\\
25. E. Chudakov et al., Report number: JLAB-THY-16-2329, arXiv [hep-ph]:1608.08686.\\
26. M.Klasen, K.Kovarik and J.Potthoff,     Phys. Rev. D {\bf95}, 094013 (2017).\\
27. Y.Hatta, Nucl.Phys.A {\bf00}, 1 (2020); O.Bruning, A. Seryi
and S. Verdu-Andres, Front.in Phys. {\bf10}, 886473 (2022).\\
28. E.Long, Technical Note (2019), T19-002.\\
29. A.Capella et al., Eur.Phys.J.C {\bf5}, 111 (1998).\\
30. M. M. Block, L. Durand and P. Ha, Phys. Rev.D{\bf 89}, 094027 (2014).\\
31. G. Soyez, arXiv [hep-ph]:0705.3672.\\
32. C.Marquet, M.R.Moldes and P.Zurita, arXiv
[hep-ph]:1702.00839.\\
33.M.A.Betemps and M.V.T.Machado, arXiv [hep-ph]:0906.5593.\\
34. A.M.Stasto, K.Golec-Biernat, J.Kwiecinski, arXiv
[hep-ph]:0007192.\\
35. N.Armesto, C.A.Salgado and Urs A.Wiedemann, Phys.Rev.Lett. {\bf94}, 022002 (2005).\\
36. N.Armesto, C.Merino, G.Parente, E.Zas, Phys.Rev.D {\bf77},
013001 (2008).\\
37. I.I.Balitsky, Nucl.Phys.B {\bf463}, 99 (1996); Phys.Rev.Lett.
{\bf81}, 2024 (1998); Phys.Lett.B {\bf518}, 235 (2001).\\
38. Y.V.Kovchegov, Phys.Rev.D {\bf60}, 034008 (1999); {\bf61},
074018 (2000).\\
39. L.N.Lipatov, Sov.J.Nucl. Phys. {\bf23}, 338 (1976);
E.A.Kuraev, L.N.Lipatov and V.S.Fadin, Sov. Phys. JETP {\bf45},
199 (1977); I.I.Balitsky and L.N.Lipatov, Sov.J.Nucl.Phys.
{\bf28}, 822 (1978).\\
40. H1 Collaboration, C.Adloff et al., Eur.Phys.J.C {\bf21}, 33
(2001); ZEUS Colaboration, J.Breitweg et al., Eur.Phys.J.C
{\bf12}, 35 (2000); S.Chekanov et al., Eur. Phys. J.C {\bf21}, 443
(2001).\\
41. J.L.Albacete, N.Armesto, J.G.Milhano, C.A.Salgado,
U.A.Wiedemann, Eur. Phys. J. C {\bf43}, 353 (2005); Phys. Rev. D
{\bf71}, 014003 (2005).\\
42. M.Krelina and J.Nemchik, Eur.Phys.J.Plus {\bf135},
444(2020).\\
43. L.Agozzino, P.Castorina and P.Colangelo, Eur.Phys.J.C {\bf74},
2828 (2014).\\
44. M.R.Adams et al. [ E665 Collaboration], Phys.Rev.Lett.
{\bf68}, 3266 (1992); Z.Phys.C {\bf67}, 403 (1995); M.Arneodo et
al. [ EMC Collaboration], Nucl.Phys.B {\bf333}, 1 (1990);
M.Arneodo et al. [ NMC Collaboration], Nucl.Phys.B {\bf441}, 12
(1995); P.Amaudruz et al. [ NMC Collaboration], Nucl.Phys.B
{\bf441}, 3 (1995).\\
45. B.Z.Kopeliovich, J.Nemchik, I.K.Potashnikova and I.Schmidt,
J.Phys.G {\bf35}, 115010 (2008); B.Z.Kopeliovich, J.Nemchik, N.N.
Nikolaev, B.G.Zakharov, Phys.Lett.B {\bf324}, 469(1994).\\
46. K.J.Golec-Biernat, M.Wusthoff, Phys.Rev.D {\bf59}, 014017
(1998).\\
47. H.Kowalski, L.Motyka, G.Watt, Phys.Rev.D {\bf74}, 074016
(2006).\\
48. B.Z.Kopeliovich, A.Schafer, A.V.Tarasov, Phys.Rev.D {\bf62},
054022 (2000).\\
49. J.Bartels, K.J.Golec-Biernat, H.Kowalski, Phys.Rev.D {\bf66},
014001 (2002).\\
50. A.H.Rezaeian, M.Siddikov, M.Van de Klundert, R.
Venugopalan, Phys.Rev.D {\bf87}, 034002 (2013).\\


\end{document}